# Artificial intelligence and the transformation of higher education institutions

**Evangelos Katsamakas**[1*], **Oleg V. Pavlov**[2], **Ryan Saklad**[2]

[1] Gabelli School of Business, Fordham University, New York, NY, U.S.A.; katsamakas@fordham.edu
[2] Worcester Polytechnic Institute, Worcester, MA, U.S.A.
* Author to whom correspondence should be addressed

## Abstract

Artificial intelligence (AI) advances and the rapid adoption of generative AI tools like ChatGPT present new opportunities and challenges for higher education. While substantial literature discusses AI in higher education, there is a lack of a systemic approach that captures a holistic view of the AI transformation of higher education institutions (HEIs). To fill this gap, this article, taking a complex systems approach, develops a causal loop diagram (CLD) to map the causal feedback mechanisms of AI transformation in a typical HEI. Our model accounts for the forces that drive the AI transformation and the consequences of the AI transformation on value creation in a typical HEI. The article identifies and analyzes several reinforcing and balancing feedback loops, showing how, motivated by AI technology advances, the HEI invests in AI to improve student learning, research, and administration. The HEI must take measures to deal with academic integrity problems and adapt to changes in available jobs due to AI, emphasizing AIcomplementary skills for its students. However, HEIs face a competitive threat and several policy traps that may lead to decline. HEI leaders need to become systems thinkers to manage the complexity of the AI transformation and benefit from the AI feedback loops while avoiding the associated pitfalls. We also discuss longterm scenarios, the notion of HEIs influencing the direction of AI, and directions for future research on AI transformation.

# 1. Introduction

The spectacular growth of generative Artificial Intelligence (AI) tools like ChatGPT since late 2022 has brought AI to the forefront of all debates about technology and its impact on the economy and society [1]. There are concerns about the future of work and the adverse social consequences of automation that may lead to a jobless future [2–4], while companies explore how to benefit from generative AI investment [5,6].

In higher education, the rapid adoption of ChatGPT brings excitement about opportunities for learning and concerns about challenges such as students cheating on their assignments [7], for instance asking ChatGPT to instantly write an essay about any topic [8]. While the initial reaction was banning generative AI, several organizations have developed guidelines about the beneficial use of such tools in higher education institutions (HEIs), such as colleges or universities. The Russel Group of universities in the UK developed five principles, emphasizing the need for "students and staff to become AI-literate," adapting "teaching and assessment to incorporate the ethical use of generative AI," upholding academic integrity and rigor, and working collaboratively to share best practices [9]. The UNESCO guidance proposes regulation of generative AI tools by government agencies and validation of the ethical and pedagogical aspects of those tools by education institutions [10].

More generally, AI brings several opportunities and challenges for teaching, learning, student support, scholarship, and administration of HEIs. AI is not a new phenomenon in education, and it has been studied for more than 30 years, as captured in several review articles [11–16] that provide a background to inform our research. Still, what is less understood is how AI will transform education [17,18] and what HEIs could do about it, especially when it comes to generative AI due to its novelty [19–22].

This article studies the AI transformation of higher education by deploying a systems approach [23,24]. We develop a causal loop diagram (CLD) model that captures the major factors that affect AI transformation in a HIE, AI advances that drive the transformation, and changes in job markets that affect graduating students. The CLD shows the feedback loop structure that defines how a HEI creates value and how AI restructures value creation in a HIE. That allows us to understand the causal mechanism underlying several AI effects relevant to HEI, such as effects on learning, academic integrity, and jobs. Our approach integrates systems thinking with economic concepts and incentives. This allows for a whole-system exploration that emphasizes the dynamic behavior of the system.

The article makes several research contributions. First, it contributes to our understanding of the AI transformation of HEIs by providing a holistic view of the driving forces and the consequences of the AI transformation. We show that investment in AI can have strategic value because AI can transform the structure of value creation in a HEI. The CLD allows us to see the strategic significance of AI within a HIE from a whole-system viewpoint, contributing to higher education economics and strategy. A key concept is the AI feedback loop, which has been studied before in other industries, such as content platforms [23], digital platforms for the common good [24], and mobile platforms [25].

Second, our CLD allows us to see the university as a complex system and derive novel insights into the complex dynamics of higher education. Our results and insights add AI transformation as a new theme to previously published system dynamics research in higher education that studied the enrollment crisis due to demographics [26], the HEI response to COVID19 [27], and tuition inflation [28].

Additionally, this article contributes to sustainability through the study of HEIs. Goal four in the United Nations Sustainable Development Goals (SDGs) [29] is about access to quality education. We show that AI can support the advancement of goal four by demonstrating that AI can help HEIs improve the quality of learning, deal with associated challenges, and better their reputation.

In addition to contributing to research, the article provides practical insights for HEI leaders seeking to understand and leverage AI in higher education. We argue that HEI leaders need to become systems thinkers to manage the complexity of the AI transformation, benefiting from the AI feedback loops while avoiding the associated pitfalls. We also aim to clarify what is new about generative AI in the broader historical context of AI use in higher education.

The following section explains the research methods. Section 3 discusses the theoretical framework. Section 4 presents the CLD model and feedback analysis. Sections 5 and 6 are the discussion and conclusions.

# 2. Method

To provide a bigpicture holistic map of the importance of AI for HEIs, we develop a causal loop diagram (CLD) of AI transformation dynamics in a HEI. Developing a CLD to map the causal feedback processes that drive the dynamic behavior of a system is widely used in many fields [27,30–32], including modeling value creation in organizations [24,25,33].

We build the CLD following an iterative process of model refinement based on relevant literature [34,35] and our study of current developments leveraging our domain expertise. We identify critical variables (factors) in the system under study. Then, we document the complex network of causal links between variables. Causal links create feedback loops, which become visible only through the modeling process. A feedback loop is reinforcing (a change in a factor amplifies via the loop) or balancing (a change is dampened via the loop). The structure and interaction of the feedback loops determine the system behavior through time. The validity of the resulting CLD model is further established by feedback from domain experts [36–38]. The theoretical framework is discussed next.

# 3. Theoretical framework

We study AI transformation in a typical HIE, emphasizing novel opportunities and challenges due to generative AI. Our theoretical framework focuses on three overlapping processes: advances in AI technology that enable the AI transformation (Section 3.1), dimensions of AI transformation in the HEI (Section 3.2), and the AI impact on the jobs for graduating students (Section 3.3).

## 3.1 Advances in Artificial Intelligence (AI) technology

With its continuous advances, AI has many promising business applications and is expected to transform our lives, business, and society [1,39–43]. Artificial intelligence as a field has a 70year history, with multiple waves of progress followed by periods of challenges called AI winters. It is a diverse field of research and practice related to creating and evaluating intelligent systems [44] with various problems (e.g., reasoning, prediction, planning, vision, language understanding), approaches, technologies, and applications (e.g. [45]). One popular approach has been creating rulebased systems that encode knowledge of experts, e.g., rules about making a medical diagnosis, but those systems have substantial limitations. Instead of capturing knowledge in software, the approach that proved most fruitful is designing algorithms that learn from data and training them with large quantities of data on powerful computers—this is the machine learning approach. Various approaches to learning are used depending on the problem: supervised learning, unsupervised learning, reinforcement learning, and others.

Most recent AI advances rely on machine learning using largescale neural networks, called deep learning, due to the multiple layers of neurons. One example is largescale neural networks for language, called Large Language Models (LLMs), that can generate text, including code, following a user prompt or a sequence of user prompts (dialogue with the user), hence generative AI. LLMs are trained using large datasets [46], and because they deal with language, they also belong to the area of AI called natural language processing (NLP). OpenAI's ChatGPT, using generalized pretrained transformer architecture with billions of parameters (weights), is the most wellknown example, amongst many, of a conversational generative AI application built on an LLM. Other generative AI applications produce images, music, videos, or multiple types of media (multimodal models), so the general term 'foundation model' is sometimes used for generative AI models. The art of writing prompts to get the best results from the system is called prompt engineering. The systems typically incorporate filters called guardrails to ensure they do not produce offensive or otherwise undesirable content. Other significant challenges and risks are discussed in subsection 3.2.5.

## 3.2 Dimensions of AI transformation in HEI

As already mentioned, the novelty of generative AI lies in the "generative" aspect of its name. Generative AI tools facilitate the creation of new content. From an economic standpoint, generative AI lowers the cost of knowledge work, especially knowledge creation tasks. Changing the economics of knowledge processes in markets and organizations can be transformative [47–50]. This is particularly true in higher education because HEIs are organizations that manage knowledge: they create new knowledge via research, deliver knowledge to students via teaching, and assess learning by asking students to do some knowledge work, for instance, write essays. In the latter, writing an essay becomes a zerocost task that takes a few seconds and does not require any familiarity with the topic. In essence, these tools provide a new interface to knowledge, akin to an advanced search engine that uses as input all information available on the web and produces as output a readytouse answer instead of sources of information.

We identify and explore five dimensions of AI transformation in HEI: student learning, academic integrity problems, faculty research productivity, administration and operations, and risks and ethics issues affecting the use of AI in HEIs.

### 3.2.1 Student learning

AI can support student learning by supporting instructors who teach or by supporting students directly. Empowering teachers and students should be the primary objective.

AI has the potential to transform teaching by supporting instructors. Instructors could use AI as a support to design programs or courses, create new education material and assignments, deliver better instruction that increases student engagement and motivation for learning, and assess learning more creatively and authentically. Faculty can also use AI to automate timeconsuming administrative tasks so that they can focus on creativity and innovation in teaching and research. AI and other Industry 4.0 technologies, such as the Internet of Things, can enable smart classrooms and the digital transformation of education management, teaching, and learning [51]. Other examples include learning analytics, educational data mining, and intelligent webbased education [11]. Cobots (collaborative robots designed to work alongside humans) assisting teachers in the classroom is another concept adapted from the manufacturing industry [52]. A largescale review of more than 4500 articles published between 20002019 [53] finds that the main research topics include intelligent tutoring systems for special education, natural language processing for language education, educational data mining for performance prediction, discourse analysis in computersupported collaborative learning, neural networks for teaching evaluation, affective computing for learner emotion detection, and recommender systems for personalized learning. Another review of 138 articles from 2016 to 2022 [12] finds five topics: assessment/evaluation, predicting, AI assistant, intelligent tutoring system, and managing student learning. AI is expected to significantly impact teaching and learning [54].

Students can use AI as a support tool to meet their learning goals via personalized adaptive learning. Applications come in various names, such as personalized learning [55], AI teaching assistants, teacherbots [56, 57], intelligent tutoring systems [58], and others. An empirical study finds that AI enhances learning through collaborative learning and an accessible research environment [59]. An experimental study in India finds that personalized technologyaided afterschool instruction improves student scores in math and language [60]. Gains attributed to the tutoring effect can be expected to be larger using more recent AI technologies, such as GPT4.

Generative AI can empower students and enhance their educational resources and experiences [61]. There are several ways that generative AI can be used in the classroom, such as a tutor, coach, or teammate [62]. Alternatively, AI can be used as a tutor or coach outside the classroom, while classroom time is used for activities that apply knowledge.

A largescale study finds that essays written by ChatGPT are of higher quality than student essays [63]. However, another experimental study finds that ChatGPT is not effective as an essay writing assistant for students [64]. Likewise, a comparison of several AI chatbots across multidisciplinary tests finds that all of them get low grades [65]. Other studies argue that instructor AIreadiness matters [66] and propose a new theoretical framework for blended learning with generative AI integration [67].

While publicly available generalpurpose tools like ChatGPT get most of the attention, the most value may come from specialized tools created with specific education objectives and trained with appropriate data or using retrieval augmented generation (RAG). An example is Khanmigo by Khan Academy (https://www.khanacademy.org/khanlabs), which aims to bring onetoone tutoring to all students and an assistant to teachers using AI. It runs on top of the OpenAI platform and is used widely as a pilot phase, but research on its efficacy is expected in 2024 [68].

### 3.2.2 Academic integrity problems

There is significant concern that ChatGPT would facilitate high levels of cheating in higher education, undermining learning and academic honesty [69,70]. However, there was already substantial cheating before ChatGPT. One study found that 15% of students had used an essay mill, an online service where students pay for people to write essays for them [71]. Another study at an Australian engineering university found that 50% of questions asked on Chegg (the largest homework help site) were answered within 90 minutes [72].

Generative AI is another instance of technology facilitating student cheating. Just two months after ChatGPT's release, an estimated onefifth to over onethird of students reported using it, with the vast majority believing they cheated using it [73]. In less than a year since the release of ChatGPT, the stock price of Chegg plunged, and the company attributed the drop to students flocking to use ChatGPT for cheating instead [74]. And as students become more familiar with the technology, they also become more effective at using it.

Moreover, academic integrity problems relate to employers seeing higher education as a signaling device [75]. When hiring, employers take many timesaving measures to reduce time and costs of the hiring process. Rather than carefully assessing each candidate's knowledge, employers will only consider applicants who graduated college and use GPA as a filter. A study found that "66 percent of employers screen candidates by grade point average (GPA), and 58 percent of employers indicated that a GPA below 3.0 all but eliminates a candidate's chances of being hired" [76]. As a result, students could perceive that graduating with a degree and GPA that employers will desire is more important than learning. This creates a strong incentive for students to cheat using AI. A study that investigated an alternative system, national college exit exams, found that they decreased a college's reputation premium by better capturing the skills of individual students [77].

In the past, there are multiple examples of instructors modifying their teaching in response to technologies making cheating easier. The calculator [71] and Wikipedia [78] were two historical examples that transformed pedagogy. In these cases, cheating was reduced by changing the type of teaching and assessment, such as calculator and noncalculator test sessions, and focusing less on rote memorization. Overall, HEIs can respond by reducing incentives to cheat, increasing the value of learning, making it harder to cheat, or increasing the risk and consequences of getting caught. However, anticheating measures have tradeoffs. For example, using online proctoring software may reduce cheating, but it also costs money, causes technological difficulties, has false positives, and reduces student's privacy. The most common initial approach by schools was AI detection software. Unfortunately, AI detection software has an extremely high false positive and false negative rate and flags the work of nonnative speakers significantly more than their peers [79].

A systematic review on cheating in online exams from 2010 to 2021 found several approaches to reduce academic dishonesty before testing [80]: strengthening student ethics, bringing the learning goal of the exams to mind, and moving away from summative assessments towards formative assessments. However, other approaches, such as randomizing questions or shifting toward essays, become less effective with widespread AI usage. Most importantly, HEIs must update their academic integrity policies, and faculty must update their course syllabi to account for generative AI. There is a need for clear policies to deal with academic integrity and plagiarism detection challenges [81]. Some courses could allow the creative use of generative AI and adjust assignments and assessment accordingly, while others could prohibit it.

### 3.2.3 Faculty research and accelerated scientific discovery

AI, such as machine learning techniques, is increasingly used in science research, and researchers are excited about its potential [82]. Still, they are also concerned about the quality of work and reproducibility of results [83]. If used correctly, generative AI can support scholarly work and faculty research productivity [84]. Such tools can support problem formulation, data collection and analysis, and writing [85]. Those include tasks such as research brainstorming, identifying research questions, hypothesis generation [86–88], summarizing or conducting a literature review, creating graphs from data, drafting parts of manuscripts and others.

However, all those uses come with challenges such as AI hallucinations (making up stuff), accuracy, completeness, quality, and others. Moreover, the ease of creating content using generative AI tools may increase academic misconduct or the mass production of lowquality papers flooding journals and the established peerreview process. Both would have significant negative consequences for scholarly publishing and research, and journals are updating their editorial policies. For instance, Science journals do not accept text written by AI tools [89]. Ultimately, the authors are responsible for all aspects of the research output, and they also need to be transparent about whether and how they used AI tools. While conversational generative AI tools have the potential to play a significant role in the research workflow, there are many open questions about the details of the practical application of those tools when it comes to automation and augmentation (table 6 in [85]), as well as a need for guidelines [90].

More broadly, AI promises to accelerate research and scientific discovery [91–93]. That promise is aligned with the knowledgecreation mission of HEIs. However, in the longer term, only large tech companies may have the computing and data resources for complex, largescale, and highimpact science research, such as Google DeepMind's AlphaFold for protein folding in biology [94] and discovering thousands of new materials in material science [95,96]. As a result, HEIs may be sidelined unless they partner with big tech companies, the research divide in higher education may get bigger, and big tech firms may become the gatekeepers of consequential research agendas.

### 3.2.4 Administration and operations: Institutional learning

Although our review of the literature on AI in higher education finds that the main focus is student learning and teaching, other HEI areas can benefit from AI [97,98]. AI and data can help improve the effectiveness and lower the operating costs of all areas in the university, such as administration of the HEI, including departments and schools; admissions to improve enrollments; academic advising to guide students and career advising [99], internships and job placement of students; alumni relations; IT, human resources, athletics, facilities, and operations [100]. Those opportunities for improvement can be seen as institutional learning. AI is about learning and institutional learning means that the HEI uses AI to become a learning organization and pursue continuous improvement while adapting to changes in its environment.

### 3.2.5 AI risks and ethics in HEIs

Generative AI has a long history [101], and while recent generative AI signifies progress, we should be aware of its limitations [102–104] and discount the hype. For instance, LLMs are probabilistic language modelers predicting how to continue the text based on patterns learned from training data. They lack causal models of understanding the world, and their outputs need critical evaluation.

ChatGPT and related tools are designed to create persuasive and authoritative output, even when they make stuff up, a wellknown problem called hallucination. This is a severe problem for education because the thing worse than not learning anything is learning the wrong things very well. AIcreated fake media, such as images and videos (deep fakes), will exacerbate the challenges to proper learning and social cohesion.

Besides clearly damaging misinformation, large quantities of poorquality content are a problem for student learning. Humans have limited time and attention (cognitive capacity), and those resources can be easily wasted in an environment where multiple services compete for user attention (attention economy) using algorithms optimized for user engagement. Moreover, poor quality content from GenAI tools may pollute the Web, affecting all users of that content, including GenAI tools that use that content for training.

Algorithmic bias is another significant concern [105]. Algorithms may reinforce decision biases when evaluating student work, admissions, job placement, etc. In a reinforcing feedback loop, bias in historical data drives algorithmic bias, which drives decision bias, which drives even more human bias and bias in the data.

AI in higher education also has a dark side related to data [106]. Data is an essential resource for AI. The need for large quantities of data creates privacy, security, and copyright risks. For instance, sensitive data about students and their behavior must be well protected. Confidential data may leak if used to interact with publicly available AI chatbots. Malicious actors can use AI for cyberattacks. Ignoring copyrights in model training is another issue, and ongoing lawsuits may affect how future generative AI systems work [107].

Training AI models often utilizes global cheap labor to label data, moderate content, or provide feedback, creating ethical concerns about labor practices [108]. Increased complexity due to fast change, loss of control, manipulation of behavior, dependence on tech firms like OpenAI controlling the AI platform, and lack of transparency and accountability are other issues due to AI that may negatively affect multiple areas of a HEI. Constant surveillance by AI [109] damages trust and meaningful education [110]. Automation itself is a risk, if not well designed, because it could have an organization do the wrong things faster and in an automated way while no one pays attention. Accountability in AImediated education practices is an issue that needs to be studied more [111]. Environmental impacts, carbon and water footprints, and energy consumption of AI data centers are also concerning [112].

All those challenges have many practical implications for how AI systems are designed and managed. Explainability, transparency, and fairness [113] of AI decisions should be important priorities in the design of AI systems. Learning analytic systems must be thoroughly audited to ensure they are fair, transparent, and robust [114]. Generative AI tools such as ChatGPT raise even more ethical challenges and call for stakeholder engagement, a systems view of benefits and risks when applications are developed, and multilevel policy interventions [115]. Human oversight and a pedagogical approach focusing on critical thinking are recommended [116]. Moreover, education on the responsible and ethical use of the new tools, and new assessment strategies are appropriate [117,118].

## 3.3 Jobs for graduating students

HEIs educate students who seek jobs after graduation. Therefore, the state of the job (labor) market and the workforce needs of companies is an essential determinant of the value of a HEI degree.

AI can be a tool that makes a worker more productive (AI augmentation) or an automation engine that eliminates the worker's job. Therefore, what jobs and how will be most impacted by AI is a complex question [119–122]. A way to approach that question is to think of a job as a set of tasks and consider how AI affects tasks. Then, a job with many tasks automated or augmented by AI will be affected the most [123,124].

Generative AI can make knowledge workers more productive. Software developers randomly assigned to use GitHub copilot, an AI coding assistant, completed their task 55% faster than the control group [125]. Moreover, using GitHub Copilot improves other important metrics, such as developer job satisfaction [126]. Collegeeducated professionals randomly assigned to use ChatGPT in a writing task took 40% less time and produced 18% higher output quality, and participants with weaker skills benefited the most [127]. Customer support workers using generative AI achieve higher productivity, but with significant heterogeneity across workers as novice and lowskilled workers benefit the most [50]. While AI can help improve the effectiveness of consultants in many tasks, there are tasks in which AI fails, implying that overreliance on AI can lower performance [128]; for instance, LLMs hallucinate and sometimes do poorly in basic math.

While individuals care about what will happen to jobs, companies care more about the optimal mix of humans and AI that maximizes the company's performance, taking into account the strengths and weaknesses of each. The interaction of companies' needs and workers' skills and preferences will determine the effect of AI on employment outcomes. For instance, a recent study using data from a large online platform finds that generative AI affects freelancers' employment and earnings negatively [129]. Our study connects job market changes due to AI with the value created by HEIs.

# 4. CLD model and analysis

We use a causal loop diagram (CLD) to map the causal mechanisms of AI transformation in a typical HEI. Our model consists of three interconnected sectors: the AI industry that drives AI advances, the focal HEI that uses AI for transformation, and the companies that offer jobs to students graduating from the HEI. A positive arrow signifies that the causeandeffect variables move in the same direction. A negative causal relationship between variables is shown as a negative arrow. Letters R and B denote Reinforcing and Balancing feedback loops.

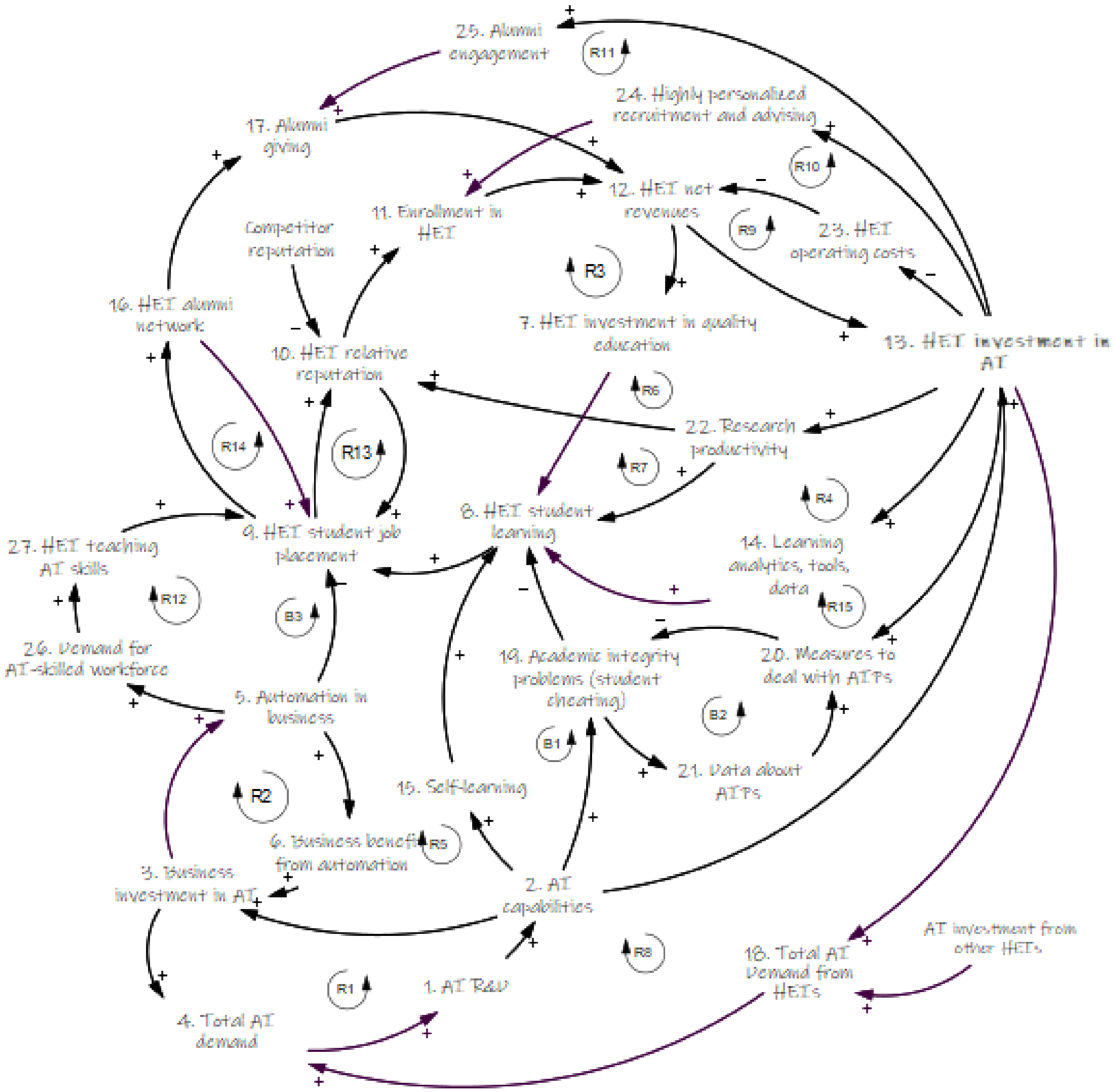

**Figure 1.** AI and the transformation of a higher education institution (HEI). *HEI investment in AI* aggregates investment for teaching, learning, research, admissions, student advising, and alumni relations. *HEI investment in quality education* aggregates all other investments in faculty, facilities, methods, advising, etc.

We identify and analyze 15 reinforcing (R) and 3 balancing (B) feedback loops that define the structure of value creation in the HEI and its interaction with the business sector and the AI industry. The feedback loops are summarized in Table 1 and discussed below.

**Table 1.** Feedback loops.

| Name of feedback loop | Variables |
|---|---|
| R1 | 1, 2, 3, 4 |
| R2 | 3, 5, 6 |
| R3 | 7, 8, 9, 10, 11, 12 |
| R4 | 13, 14, 8, 9, 10, 11, 12 |
| R5 | 2, 15, 8, 9, 10, 11, 12, 13, 18, 4, 1 |
| R6 | 22, 10, 11, 12, 13 |
| R7 | 22, 8, 9, 10, 11, 12, 13 |
| R8 | 2, 13, 18, 4, 1 |
| R9 | 23, 12, 13 |
| R10 | 24, 11, 12, 13 |
| R11 | 13, 25, 17, 12 |
| R12 | 26, 27, 9, 10, 11, 12, 13, 18, 4, 5 |
| R13 | 9, 10 |
| R14 | 9, 16 |
| R15 | 20, 19, 8, 9, 10, 11, 12, 13 |
| B1 | 2, 19, 8, 9, 10, 11, 12, 13, 18, 4, 1 |
| B2 | 19, 21, 20 |
| B3 | 5, 9, 10, 11, 12, 13, 18, 4 |

## 4.1 Advances in AI technology

We start with feedback loops related to AI advances (improved AI capabilities) in the AI industry because AI advances create opportunities for AI transformation in HEIs. An example is the development of LLMs discussed previously.

R1: Due to AI research (R&D), AI capabilities improve and encourage more business investment in AI, motivating the AI industry to invest even more in AI R&D.

R2: As AI capabilities improve, businesses invest more in AI, thus increasing business automation. More automation benefits business and encourages even more investment in AI.

Loops R1 and R2, supported by R5 and R8 discussed below, are the primary economic forces driving AI advances. We focus on the AI transformation within the HEI next.

## 4.2 Student learning

The following loop, R3, is the most fundamental feedback process that creates value for students and financially sustains a typical college or a university: The HEI prospers by investing in quality education because it improves student learning and student job placement affects positively the HEI reputation, which ensures enrollment and revenue.

The following loop captures the AI contribution to student learning. R4: The AI investment leads to better learning analytics, AI tools, and data, which improves student learning, allows students to find good jobs, and builds HEI's reputation. A strong reputation contributes to healthy enrollments and revenues that allow more investment.

Another loop that affects student learning is R5: Advances in AI capabilities facilitate students' selflearning, which adds to overall student learning.

A tradeoff between formal learning and selflearning is apparent here. Suppose students do an increasing amount of their learning through selflearning. In that case, the position of the HEI is weakened over time because fewer students will be interested in enrolling, or those enrolled will be asking for tuition discounts.

## 4.3 Student academic integrity problems

The HEI's business model might be undermined by the AIassisted integrity problems, as captured by loop B1: Better AI leads to more academic integrity problems (AIPs), such as student cheating, which negatively affects student learning, job placement, and the HEI reputation.

The HEI can use data about AIPs and AI to fight academic integrity problems, as shown in loop B2. When AIPs are low, measures to deal with AIPs will be low. As AIPs increase, the HEI will increase its efforts to deal with AIPs.

It is in the HEI's interests to invest in measures to deal with AIPs, as shown in R15.

## 4.4 Faculty research

R6: AI investment supporting the research productivity of the HEI faculty has a positive effect on the reputation of the HEI and leads to more robust enrollment numbers and positive net revenue.

R7: AI investment supporting the research productivity of the HEI faculty adds value to student learning due to researchteaching complementarity.

## 4.5 HEI administration and operations

R8: Advances in AI motivate the HEI to invest more in AI.

R9: The HEI uses AI to lower operating costs, so there is a higher net revenue for investments in quality education and AI supporting it.

R10: The HEI uses AI to support admissions and improve new enrollment numbers, student support, student retention, and graduation rates, thus increasing total enrollment in HEI.

R11: The HEI uses AI to support alumni engagement and improve alumni giving.

## 4.6 Job placement

R12: Business adoption of AI is an opportunity for job placement of students who acquire AIcomplementary skills. Those skills are discussed in more detail later.

The jobsubstitution effect of AI manifests itself as a balancing loop B3: Business automation is a challenge for job placement because it lowers the number of available jobs.

R13: The HEI relative reputation and the student job placement reinforce each other.

R14: When the HEI does well in student job placement, it enlarges its alumni network, which is an opportunity for larger alumni giving (that helps all the other investments), but also improves job placement of new graduates.

## 4.7 Additional feedback analysis

AI helps the HEI improve the quality of its offered services (R4, R6, R10, R11) and lower the cost of operations for a given level of service (R9). Those feedback loops together work for the benefit of a wellmanaged HEI. As long as AI keeps advancing, driven primarily by business demand, the reinforcing feedback loops create a virtuous cycle for a HEI that improves its reputation relative to its competitors. However, those same loops will hinder the HEI that falls behind in the competitive higher education market because HEIs compete on reputation. In that context, AI investments can help a HEI differentiate itself and soften competition. In addition, measures to fight AIPs can differentiate a HEI if AIPs become a significant problem in the higher education sector.

Data is a valuable resource for the effective use of AI in HEIs (see, for instance, R4 and B2). Indeed, the more data the HEI collects about all areas (learning effectiveness, job placement, alumni, reputation, admissions, student retention, etc.), its AI will be more effective. For a HEI, value comes from AI plus data. Therefore, data creates additional essential feedback loops not shown in Figure 1 to keep the CLD concise. Those AI feedback loops are similar to the AI feedback loops captured in [23] in a digital platforms setting.

# 5. Discussion

We discuss research insights and theoretical contributions, implications for academic leadership and policymakers, and future research directions.

## 5.1 Findings and research implications

This article takes a novel complex systems approach to how a HEI creates value and how AI affects those valuecreation processes. The article explores the effects of AI in higher education using a CLD, and it identifies multiple feedback loops and their interactions.

Although AI has been relevant in higher education for the past twenty years, new waves of AI, such as the generative AI today, create opportunities for transformation in a HEI. AI can help a HEI improve learning and increase its reputation, student enrollment, and revenue through several reinforcing loops. At the same time, current AI advances intensify academic integrity problems, a balancing loop, and if not adequately addressed, may undermine learning and the associated benefits for HEIs.

If we focus on academic integrity problems, then a potential danger is education turning into a ‘market for lemons’ in the eyes of employers if HEIs do not effectively address those problems. With increased student cheating, the employment market for graduates becomes a ‘lemon market’, as employers cannot easily discern which students learned. In extreme cases, the employment market collapses. Moreover, if a HEI has AIPs that are not in line with the peers, the institution puts itself at a competitive disadvantage by not investing in efforts to deal with AIPs. In other words, if a college has very high cheating rates, it will not be able to compete. Therefore, a HEI must maintain the proper amount of anticheating measures to be better or on par with competitors. This dynamic can also be interpreted as a ‘tragedy of the commons’, with the HEI’s reputation as a common resource and students acting as selfinterested parties; as the regulator of the common, the HEI must develop measures to fight academic integrity problems or risk that its reputation would collapse.

Moreover, students who graduate from HEIs expect to find jobs, so job placement is a crucial factor in the system under study. The CLD shows the crucial role of a HEI’s student job placement because it affects enrollments and revenues through several pathways. Job placement depends on student learning, a HEI's relative reputation, and job availability. AI impacts all three factors through several pathways, as shown in Figure 1. Therefore, the HEI needs to make the best use of AI to prepare its students for a job market shaped by AI, while other HEIs are likely to do the same, creating new AI opportunities and challenges over time.

In the business world, AI automation lowers the demand for labor but increases the demand for new skills. Successful HEIs adapt to those changes by teaching AI complementary skills. In the longterm scenario where AI automates all or most of the jobs, the current model of HEI collapses (see feedback loops R3, R4). HEIs, as we know them today, may disappear if there will be no demand for degrees, perhaps except for a small number of elite HEIs educating the government and business leaders. Those HEIs that survive and thrive will need models disconnected from degrees for jobs. They will need to create value in other ways, perhaps teaching humans leisure skills, providing lifelong learning training to humans (instead of intensive higher ed degrees as we know them today), or training and tuning AI systems in partnership with companies. If humans are supported by some kind of universal basic income (UBI) [130] due to the lack of jobs, then part of that income could be support for lifelong learning, a universal basic lifelong learning income (UBLI). Under this scenario, government support will be a source of revenue for the future HEIs.

An alternative longterm scenario is that AI will become a new platform for new types of jobs, and there will be an enormous demand for people to fill those jobs (similar to jobs in factories after the industrial revolution or office jobs with the adoption of computing). In that case, the future of HEIs is bright, especially if the job market is very fluid and people need multiple degrees over their lifetime.

A HEI can grow and prosper or decline, depending on its AI investment and policies. Our article shows that AI rewires the feedback loop structure that defines how a HEI creates value. Therefore, AI feedback loops can play an essential role in a HEI. This insight adds to previous research that finds that AI feedback loops play an important role in digital platforms [23,25]. Moreover, our work highlights that AI transformation is at the forefront of the ongoing digital transformation of higher education [131–134].

## 5.2 Lessons for academic leadership

AI advances in the form of generative AI create several opportunities for AI transformation, including the promise to bring HEIs closer to the vision of personalized AI assistants that support students, faculty, and administrators. In that context, our research provides a first map of AI causal mechanisms to help HEI leaders navigate an uncharted landscape of opportunities and pitfalls.

Leaders can use the CLD to build intuition and evaluate the benefits and risks of various scenarios and HEI policies. Our discussion of feedback loops in section 4 is a starting point in that direction, but many other policies can be evaluated. For instance, a policy focused on costcutting at the expense of education quality risks placing the HEI at a reinforcing decline trajectory primarily due to the feedback loop R3. If AI is used to support such a policy, then AI will speed up the decline, whereby revenues keep getting lower, and the HEI keeps costcutting until both approach zero.

A crucial question for academic leaders is what competencies and skills will students need to find a job. Following our earlier exploration, students should avoid competing headtohead with AI. Instead, they need foundational human skills that AI lacks, for instance, critical thinking, planning, complex problemsolving, creativity, lifelong learning, communication, management, and collaboration. Students need to learn how to learn and think in ways that differentiates them from machine learning. If AI becomes ubiquitous in firms, humans need skills that complement what AI can do well to benefit from AI. That includes skills to build, train, deploy, use, and manage AI systems, identify valuable use cases, devise AI strategies, lead teams or companies, etc. Moreover, students need to acquire those AI complementary skills in a way (quality, breadth, and depth) that allows them to compete effectively against other humans seeking similar jobs. For instance, managers that use AI effectively may replace those that do not.

HEIs need to monitor changes in the job market [3] and remain adaptive. For instance, a recent study argues that LLMs can transform the role of a data scientist from coding and datawrangling to assessing and managing analyses performed by AI tools [135]. In that case, skills related to strategic planning, coordinating resources, and overseeing the product life cycle become more important, and teaching data scientists must adapt accordingly, perhaps gradually over a period of time.

The AI effects on productivity and automation are also relevant to what happens to jobs within HEIs. Will AI make instructors, administrators, and staff more productive and their jobs more fulfilling? Will AI replace instructors, administrators, and staff in the longer term? Multiple effects play a role simultaneously, and the specified time horizon matters. However, a crucial framing question is: what does the HEI want to achieve with AI? The university policy and mission matters. For instance, a university that does not grow and does not aspire to the highest learning standards may manage with a small number of instructors, administrators, and staff, provided all those roles become more productive and many tasks are automated. However, a studentcentered and humancentered university that appreciates its people and wants to make their jobs more fulfilling may be successful by providing a superior education and differentiating itself from competitors that focus on costcutting.

A related issue is the future direction of AI. Our exploration suggests that the direction of AI advances is not predefined [136], and the social responsibility of a university lies in prioritizing how AI can empower humans by augmenting jobs rather than eliminating them [137]. As a starting point, HEIs could focus on designing and adopting personalized AI assistants for higher education: for faculty, students, staff, administrators (including department chairs and deans), advising, and more. At the same time, there is a need for careful integration of generative AI tools into education [138]; during the COVID19 pandemic, students suffered both academically and socially, and we relearned that education is a "deeply human act rooted in social interaction" (p. 7). Beyond the boundaries of the education sector, HEIs could promote AI assistants for various roles (e.g., financial analyst, CEO) across all industries and teach students accordingly.

In that direction, our CLD suggests that a single HEI has very little influence over the direction of AI, but a consortium of HEIs can have a meaningful influence. Moreover, similar to the proposals in the healthcare industry [139], there is value in opensource LLMs developed by a consortium of HEIs. Those insights suggest a tradeoff for a HEI: Investment in AI is a tool for getting ahead of its competition, but if it wants to influence the direction of AI meaningfully, the HEI needs to collaborate with other HEIs. Along those lines, AI advances could support education research that provides novel, rigorously validated insights into teaching and learning methods that could benefit all HEIs.

Overall, AI promises several benefits but entails challenges, and ultimately, it depends on what policy the HEI wants to follow and how to position itself by leveraging AIenabled transformation while protecting itself from the associated pitfalls. Regarding generative AI, HEIs deal with fastchanging technology and applications. Therefore, HEIs need to be adaptive. Start with smallscale experiments by faculty, students and staff, then learn from that, aggregate the experiences and perceptions, allow for more stability, and then plan and develop more comprehensive policies and guidelines. Leaders must take a balanced and realistic approach to their assessments and proceed cautiously. At this point, both businesses and HEIs are exploring how to take advantage of the latest AI innovations. Generative AI is the current novel tech, and it is natural to be overhyped and accompanied by an aura that will solve all the problems. This pattern is typical in technology space and tends to appear every few years. Contrary to conventional marketing, AI cannot solve all the problems but can create many new benefits and challenges. As long as AI advances at a fast pace, HEIs and AI will coevolve. In that coevolutionary process, universities could also learn from partnering with AI firms or other universities.

All the complexities associated with the rapid AI adoption underscore the need for academic leaders who are system thinkers. They must study the feedback loops that define the structure of value creation and determine the system behavior. Moreover, new technologies such as AI can bring a significant restructuring by creating new feedback loops, rewiring existing ones, and strengthening or weakening others. Leaders should aim to leverage those feedback loops for their benefit. A systems approach appreciates complexity, takes a wholesystem view, understands that system behavior over time is often nontrivial and counterintuitive, and considers the unintended consequences. For instance, an overreliance on costcutting approaches can place a HEI into a selfreinforcing decline. Another underappreciated systemic risk arises from the uniform adoption of identical AI models and practices across all HEIs, leading to an escalation in academic competition.

## 5.3 Limitations and future research directions

This article provides the first holistic map of AI transformation in HEIs. Future work could enhance and refine that map or go deeper into specific aspects of the map. While the level of analysis here is a HEI, future research could be more micro, taking an indepth look into specific university functions, or more macro, using the higher education sector as a unit of analysis.

At the sector level, 'superstar effects' may be significant in the longer term. A global education marketplace and ubiquitous online access create positive feedback loops where the positive reputation of a school, program, course, or instructor keeps increasing. As a result, superstars may emerge similar to superstars in sports or entertainment industries.

Our model suggests that the AI industry plays a significant role because it drives AI advances affecting businesses and HEIs. More work is needed on how established and startup tech and edtech companies affect the broader transformation of the higher education sector. More generally, higher education has a lot to learn from other sectors, such as media and advertising, already transformed by AI and related digital technologies, and this has to be a topic of rigorous future research.

Future research could study various scenarios or interventions in more detail. For instance, potential decreases or a plateau in AI capabilities through regulations, limitations of current AI approaches, another AI winter, black swan events, or otherwise, could cause significant economic shocks to HEIs and businesses. Approaches to prevent 'lemon market' effects on graduates, including exit exams, microcertifications, and employment tests, should be examined. Future educational advances, like customized courses and AI tutoring, will need to be studied empirically.

Methodologically, the current article focuses on a CLD, or qualitative system dynamics. A natural next step is developing and analyzing analytical or computational models [28,140] to derive additional insight into AI in higher education.

Lastly, this article explores AI effects on a typical HEI today [26]. Future research needs to consider and evaluate novel business models for higher education.

# 6. Concluding remarks

This article presents the first causal loop diagram of the AI transformation in HE, providing a holistic view of how important variables interact to drive AI investment and impact. We show that several reinforcing and balancing AI feedback loops work together to impact value creation in a HEI that interacts with companies that provide jobs for students, and the AI industry that drives AI advances. The model shows that the HEI invests in AI to improve teaching, research, and administration, but it must also adapt to student job market changes and take measures to deal with academic integrity problems. Student job placement is a crucial factor for the sustainability of the HEI model. Therefore, the HEI needs to emphasize AI complementary skills for its students. However, HEIs face a competitive threat and several traps that may lead to a decline. For instance, HEI policies focusing on excessive costcutting may reinforce its decline. In the long term, the current HEI model will not be viable if AI automation in companies becomes increasingly labordisplacing.

The article makes several contributions. It provides a systemic view of AI in education and proposes that academic leaders should become system thinkers to benefit from AI opportunities. It contributes to our understanding of the AI transformation of higher education from a complex systems perspective that focuses on the etiology and the consequences of AItransformed value creation in HEIs. The article integrates systems thinking and economic concepts to add to higher education economics and strategy with an emphasis on dynamic complexity. Moreover, it contributes to our thinking of how AI can support the sustainability of HEIs and highquality education, which is one of the UN's sustainable development goals. Another significant contribution is connecting the HEI model affected by AI with job market factors, also affected by AI. Still, a complex systems view of higher education suggests that we are just starting to explore the impact of AI on that sector. Therefore, the article outlines several directions for future research on AI transformation and provides a basis for developing quantitative models.